\documentclass[12pt,preprint]{aastex}

\usepackage{graphics}

\newcommand{\be}{\begin{equation}}
\newcommand{\ee}{\end{equation}}
\newcommand{\bd}{\begin{displaymath}}
\newcommand{\ed}{\end{displaymath}}

\slugcomment{to appear in ApJ}

\shorttitle{Jet formation in BL Lacertae objects}
\shortauthors{Cao X}

\begin{document}

\title{Jet formation in BL Lacertae objects with different accretion
modes}

\author{Xinwu Cao}
\affil{Shanghai Astronomical Observatory, Chinese Academy of Sciences,
80 Nandan Road, Shanghai, 200030, China\\
Email: cxw@center.shao.ac.cn}

\clearpage

\begin{abstract}

We estimate the masses of massive black holes in BL Lac objects
from their host galaxy luminosity. The power of jets and central
optical ionizing luminosity for a sample of BL Lac objects are
derived from their extended radio emission and the narrow-line
emission, respectively. The maximal jet power can be extracted
from a standard thin accretion disk/spinning black hole is
calculated as a function of dimensionless accretion rate $\dot{m}$
($\dot{m}=\dot{M}/\dot{M}_{\rm Edd}$). Comparing with the derived
jet power, we find that the accretion disks in most BL Lac objects
should not be standard accretion disks. For a pure advection
dominated accretion flow (ADAF), there is an upper limit on its
optical continuum luminosity due to the existence of an upper
limit $\dot{m}_{\rm crit}$ on the accretion rate. It is found that
a pure ADAF is too faint to produce the optical ionizing
luminosity of BL Lac objects derived from their narrow-line
luminosity. We propose that an ADAF is present in the inner region
of the disk and it becomes a standard thin disk in the outer
region in most BL Lac objects, i.e., ADAF$+$SD(standard disk)
scenario. This ADAF$+$SD scenario can explain both the jet power
and optical ionizing continuum emission of these BL Lac objects.
The inferred transition radii between the inner ADAF and outer SD
are in the range of $40-150~GM_{\rm bh}/c^2$, if the disks are
accreting at the rate $\dot{m}=0.01$.

\end{abstract}

\keywords{galaxies: active---BL Lacertae objects:
general---galaxies: jets---accretion, accretion disks---black hole
physics}

\section{Introduction}

Most BL Lac objects have featureless optical and ultraviolet
continuum spectra, and only a small fraction of BL Lac objects
show very weak broad emission lines \citep{vv00}. The
low-energy-peaked BL Lac(LBL) objects exhibit similar radio and
optical continua as flat-spectrum radio quasars (FSRQ)
\citep{up95}. Similar to other quasars, FSRQs have strong
broad-line emission. The broad emission lines of quasars are
produced by the optical/UV continua reprocessed by distant gas
clouds in broad-line regions (BLRs). The optical/UV continuum
emission ionizing the gases in BLRs is believed to be from the
accretion disks surrounding massive black holes \citep{sm89}. Low
accretion rates $\dot{m}$ may lead the accretion flows to be
advection dominated \citep{ny95}. The emission from an ADAF is
very faint due to its low radiative efficiency and low accretion
rate. Also, low gas densities are expected in their environment
for low accretion rates $\dot m$. These can account for the
feature of no emission lines (or very weak) in BL Lac objects
naturally if their accretion disks are accreting at low rates
$\dot {m}$. For some individual BL Lac objects, it is found that
the accretion rates are indeed very low, for example, $L_{\rm
bol}/L_{\rm Edd}\sim 10^{-4}$ for Mkn501 \citep{b02}. \citet{cd02}
proposed that advection dominated accretion flows (ADAFs) may be
present in most BL Lac objects. They suggested a possible
evolutionary sequence for blazars. The detailed calculations on
the spectra of the ADAFs indicate that most BL Lac objects may
have ADAFs surrounding their massive black holes \citep{cao02a}.
On the other hand, FR I radio galaxies are believed to be the
misaligned BL Lac objects in the frame of unification schemes of
active galactic nuclei (AGNs) (see Urry \& Padovani 1995 for a
review). \citet{c99} estimated the upper limits of the disk
emission of FR I radio galaxies from the observed optical core
fluxes and found that the optical core luminosity of FR I radio
galaxies is far lower than the Eddington one $L_{\rm Edd}$,
typically $\lambda L_{\rm c,opt}/L_{\rm Edd}\le 10^{-4}$, if the
black holes in FR I radio galaxies are around 10$^9$~$M_\odot$.
\citet{gc01} used the optical luminosity of the host galaxy to
estimate the central black hole mass of FR I radio galaxy. The
central ionizing luminosity is estimated from the power of jets in
FR I galaxies. They found that almost all FR I radio galaxies are
accreting at very low rates: $\dot{m}\le 0.001-0.01$.

Relativistic jets have been observed in many BL Lac objects and
FSRQs, which are believed to be formed very close to the black
holes. The spin energy of a black hole might be extracted to power
the jet by magnetic fields supported by a surrounding accretion
disk (the Blandford-Znajek mechanism; Blandford \& Znajek 1977).
This process has been widely regarded as the major mechanism that
powers radio jets in AGNs \citep{bbr84,r82,wc95,ms96}. However,
\citet{ga97} doubted the importance of the Blandford-Znajek
process. For a black hole of a given mass and angular momentum,
the strength of the Blandford-Znajek process depends crucially on
the strength of the poloidal field threading the horizon of the
hole. The magnetic field threading a hole should be maintained by
the currents situated in the inner region of the surrounding
accretion disk. They argued that the strength of the field
threading a black hole has been overestimated. Livio, Ogilivie, \&
Pringle (1999; hereafter L99) re-investigated the problem and
pointed out that even the calculations of \citet{ga97} have
overestimated the power of the Blandford-Znajek process, since
they have overestimated the strength of the large-scale field
threading the inner region of an accretion disk, and then the
power of the Blandford-Znajek process.  The strength of the
large-scale field scales with the disk thickness, and it is very
weak if the field created by dynamo processes for thin disk cases.
L99 estimated the maximal jet power extracted from an accretion
disk on the assumption that the toroidal field component is of the
same order of the poloidal field component at the disk surface.
They argued that the maximal jet power extracted from an accretion
disk (the Blandford-Payne mechanism; Blandford \& Payne 1982)
dominates over the maximal power extracted by the Blandford-Znajek
process (L99). Apart from the strength of the field threading a
disk, the acceleration of the jet is also governed by the magnetic
field configuration and the structure of the disk
\citep{s91,ol98,ol01,cs02}. It is unclear how much power can
actually be extracted from a magnetized accretion disk without
considering its large-scale field configuration \citep{cao02b}.

In this paper, we use a sample of low-energy-peaked BL Lac objects
to explore the jet formation in BL Lac objects.
The cosmological parameters $H_{0}=50$ kms$^{-1}$ Mpc$^{-1}$ and
$q_{0}=$0.5 have been adopted in this work.

\section{Sample}

\citet{u00} used the Hubble Space Telescope (HST) WFPC2 camera to
survey 132 BL Lac objects. The host galaxies of 80 BL Lac objects
are detected, among which there are 29 LBLs. We use these 29
sources as our sample to study the jet formation in BL Lac
objects. We search the literature and find extended radio emission
data of these sources except 0454$+$844. We use the luminosity of
the narrow-line {[O\,{\sc ii}]} at 3727~$\rm \AA$ to estimate the
optical ionizing luminosity. For those the emission data of
{[O\,{\sc ii}]} being unavailable, we estimate the {[O\,{\sc ii}]}
emission from other emission lines adopting the line ratio
proposed by \citet{f91}.  The narrow-line emission data is
available for 22 sources in this sample. The data are listed in
Table 1.

\section{Black hole mass}

In order to estimate the central black hole masses of BL Lac objects,
we use the relation between host galaxy absolute magnitude $M_R$
at $R$-band and black hole mass $M_{\rm bh}$ proposed by
\citet{md02}
 \be
 \log_{10}(M_{\rm bh}/M_{\odot})=-0.50(\pm0.02)M_R -2.96(\pm0.48).
 \label{mrmbh}
\ee

The black hole masses of some sources in our sample have already been estimated from
their stellar dispersion velocity by using $M_{\rm bh}-\sigma$ relation
\citep{fm00,g00}. The estimated black hole masses from the stellar
dispersion velocity are:
$\log(M_{\rm bh}/M_{\odot})=8.65$(0521$-$365), 8.51(1807$+$698), and
8.10(2201$+$044) \citep{b03}.  \citet{f02} estimated the black hole masses
in the same way: $\log(M_{\rm bh}/M_{\odot})=$8.90(1807$+$698), and
8.27(2201$+$044). These results agree fairly well with the black hole
masses  estimated in this paper from the host galaxy luminosity:
$\log(M_{\rm bh}/M_{\odot})=8.68$(0521$-$365), 9.00(1807$+$698), and
8.32(2201$+$044) (see Table 1).

\section{Jet power}

The relation  between jet power and radio luminosity proposed by
\citet{w99} is \be Q_{\rm jet}\simeq 3\times 10^{38}f^{3/2}L_{\rm
ext,151} ^{6/7}~ {\rm W}, \label{qjetrad} \ee where $L_{\rm
ext,151}$ is the extended radio luminosity at 151 MHz in units of
10$^{28}$ W~Hz$^{-1}$~sr$^{-1}$. \citet{w99} have argued that the
normalization is very uncertain and introduced the factor $f$ to
account for these uncertainties. They use a wide variety of
arguments to suggest that  $1\leq f\leq 20$. In this paper, we
adopted the low limit $f=1$ in most cases. This relation was
proposed for FR II radio galaxies and quasars. We adopt this
relation to estimate the power of jets in BL Lac objects which is
believed to be a good approximation, since BL Lac objects have
similar radio properties as radio quasars. For most BL Lac
objects, their radio/optical continuum emission is strongly beamed
to us due to their relativistic jets and small viewing angles of
the jets with respect to the line of sight (e.g. Fan 2003). The
low-frequency radio emission (e.g. 151 MHz) may still be Doppler
beamed. We therefore use the extended radio emission detected by
VLA to estimate the jet power. The observed extended radio
emission is $K$-corrected to 151 MHz in the rest frame of the
source assuming $\alpha_{\rm e}=0.8$ ($f_{\nu}\propto
\nu^{-\alpha_{\rm e}}$) \citep{c99}.

\section{Jet formation mechanisms}

L99 have calculated the maximal jet power can be extracted from a rapidly
spinning black hole/magnetized accretion disk. Our calculations
mainly follow L99's spirit, but in more detail.

\subsection{Thin disks}

The maximal power of the jet accelerated by an magnetized
accretion disk is

\be L_{\rm BP}^{\rm max} =4\pi \int {\frac {B_{\rm pd}^2}{4\pi}}
R^2\Omega(R) dR, \label{lbpsd}\ee
where $B_{\rm pd}\sim
B_{\varphi}$ is assumed, and $B_{\rm pd}$ is the strength of the
large-scale ordered field at the disk surface.

The origin of the ordered field that is assumed to thread the disk
is still unclear.  An instability may arise if a strong jet is accelerated
by the large-scale field of the disk advected in by the accretion flow
\citep{l94,cs02}.
The strength of the field at the disk surface is
usually assumed to scale with the pressure of the disk, as done in
\citet{ga97}. However, L99 pointed out that the large-scale field
can be produced from the small-scale field created by dynamo
processes as $B(\lambda)\propto \lambda^{-1}$ for the idealized
case, where $\lambda$ is the length scale of the field
\citep{tp96,r98}. So, the large-scale field threading the disk is
related with the field produced by dynamo processes approximately
by (L99)

\be B_{\rm pd}\sim {\frac {H}{R}} B_{\rm dynamo}. \label{mfpd}\ee

The scale-height of the disk $H/R$ is given by
\citep{ln89}

\be {\frac {H}{R} } =15.0\dot{m}r^{-1}c_{2}, \label{hr}\ee where
the coefficient $c_{2}$ is defined in Novikov \& Thorne (1973,
hereafter NT73), and the dimensionless quantities are  defined by

\bd r={\frac {R}{R_{\rm G}}},~~~ R_{\rm G}={\frac {GM_{\rm
bh}}{c^2} },~~~ {\dot m}={\frac {\dot M}{\dot M_{\rm Edd}}},~~~
\ed and \be \dot{M}_{\rm Edd}={\frac {L_{\rm Edd}}{\eta_{\rm
eff}c^2}}=1.39\times 10^{15} m~~{\rm kg~s^{-1}},~~~ m={\frac
{M_{\rm bh}}{M_\odot}} \ee where $\eta_{\rm eff}=0.1$ is adopted.

The dimensionless scale-height of a disk $H/R$ is in principle a
function of $R$, and it reaches a maximal value in the inner
region of the disk (Laor \& Netzer 1989). We adopt the maximal
value of $H/R$ in the estimate of large-scale field strength
$B_{\rm pd}$ at the disk surface as done by \citet{cao02b}.

As L99, the strength of the magnetic field produced by dynamo
processes in the disk is given by

\be
{\frac {B_{\rm dynamo}^2}{4\pi}} \sim
{\frac {W}{2 H}}, \label{mfdyn0}
\ee
where $W$ is the integrated shear stress of the disk, and $H$ is the
scale-height of the disk. For a relativistic accretion disk, the
integrated shear stress is given by Eq. (5.6.14a)
in NT73. Equation (\ref{mfdyn0}) can be re-written as

\be
B_{\rm dynamo}=3.56\times 10^8 r^{-3/4}m^{-1/2}
A^{-1}BE^{1/2} {\rm gauss}, \label{mfdyn}
\ee
where $A$, $B$ and $E$ are general relativistic correction factors
defined in NT73.

In standard accretion disk models, the angular velocity of the matter
in the disk is usually very close to Keplerian velocity. For a
relativistic accretion disk surrounding a rotating black hole,
the Keplerian angular velocity is given by

\be
\Omega(r)
=2.034\times 10^5{\frac {1}{m(r^{3/2}+a)} }~{\rm s}^{-1}, \label{angv}
\ee
where $a$ is dimensionless specific angular momentum of a rotating
black hole.

We use Eqs. (\ref{mfpd})$-$(\ref{angv}), the maximal power of the jet
accelerated from a magnetized disk is available by integrating
Eq. (\ref{lbpsd}), if some parameters: $m$, $\dot{m}$, $a$,
are specified.

As discussed in L99, the power extracted from a rotating black
hole by the Blandford-Znajek process is determined by the hole
mass $m$, the spin of the hole $a$, and the strength of the
poloidal field threading the horizon of a rotating hole $B_{\rm
ph}$:

\be
L_{\rm BZ}^{\rm max} = {1 \over 32} \omega_{\rm F}^2 B_\bot^2 R_{\rm h}^2 c
 a^2, \label{lbzsd}
\ee for a black hole of mass $m$ and dimensionless angular
momentum $a$, with a magnetic field $B_\bot$ normal to the horizon
at $R_{\rm h}$. Here the factor $\omega_{\rm F}^2 \equiv
\Omega_{\rm F} (\Omega_{\rm h} - \Omega_{\rm F}) / \Omega_{\rm
h}^2$ depends on the angular velocity of field lines $\Omega_{\rm
F}$ relative to that of the hole, $\Omega_{\rm h}$. In order to
estimate the maximal power extracted from a spinning black hole,
we adopt $\omega_{\rm F} = 1/2$. As the field $B_\bot$ is
maintained by the currents in the accretion disk surrounding the
hole, the strength of $B_\bot$ should be of the same order of that
in the inner edge of the disk, and $B_\bot\simeq B_{\rm pd}(r_{\rm
in})$ is therefore adopted.

\subsection{Advection dominated accretion flows}

In the case of an ADAF surrounding a rapidly spinning black hole, the
maximal jet power extracted by the Blandford-Znajek mechanism was
calculated by \citet{an99}. Here, we calculate the jet power in a
similar way as done in the last sub-section.

The pressure of an ADAF is given by \cite{ny95} \be p=1.71\times
10^{7}\alpha^{-1}c_1^{-1}c_3^{1/2}m^{-1}\dot m ~r^{-5/2} {\rm
N~m^{-2}}, \label{pressure} \ee where $\alpha$ is the viscosity
parameter, $c_1$ and $c_3$ are given in \citet{ny95} \be
c_1={\frac {(5+2\epsilon^{\prime})}{3\alpha^2}}g(\alpha,
~\epsilon^{\prime}) \ee and \be c_3={\frac
{2(5+2\epsilon^{\prime})}{9\alpha^2}}g(\alpha,
~\epsilon^{\prime}). \ee Two parameters $\epsilon^\prime$ and
$g(\alpha,~\epsilon^\prime)$ are \be \epsilon^{\prime}={\frac
{1}{f_{\rm adv}} } \left( {\frac {5/3-\gamma}{\gamma-1}}\right)
\ee and \be g(\alpha,~\epsilon^{\prime})=\left[1+{\frac
{18\alpha^2}{(5+2\epsilon^{\prime})^2}} \right]^{1/2}-1, \ee where
the parameter $f_{\rm adv}$, which lies in the range 0$-$1, is the
fraction of viscously dissipated energy which is advected;
$\gamma$ is the ratio of specific heats. So, the value of the
parameter $\epsilon^\prime$ is in the range of $0-1$. As done by
\citet{an99}, we assume $p_{\rm mag}\sim~\alpha p$, and two
parameters $\alpha=1$ and $\epsilon^\prime=1$ are adopted to
maximize the pressure (see Cao \& Rawlings 2003 for detailed
calculations). As the accretion rate of an ADAF should be less
than the critical one $\dot m_{\rm crit}$, we can calculate the
maximal jet power extracted from a spinning black hole of mass $m$
and dimensionless angular momentum $a$ from Eq. (\ref{lbzsd})
assuming $B_\bot\simeq~B$, if $\dot m=\dot m_{\rm crit}$ is
substituted into Eq. (\ref{pressure}). The results of the extreme
case for $a=1$ have already been given in \citet{an99}.

\section{Spectra of the disks}

For normal bright AGNs, the bolometric luminosity can be
estimated from their optical luminosity  by \citep{k00} \be L_{\rm
bol}\simeq 9\lambda L_{\rm \lambda,opt}, \label{lbolopt}\ee which
may not be valid for the cases of ADAFs.

For BL Lac objects, the observed optical continuum emission may be
dominated by the beamed synchrotron emission from the relativistic
jets. We use narrow-line emission to estimate the ionizing
luminosity that is believed to be photo-ionized by the disk
emission \citep{rs91}. The ionizing luminosity of BL Lac objects
can also be derived from their broad-line emission (e.g., Cao
2002a, Wang, Staubert, \& Ho, 2002). Here we use the equivalent
width of the line {[O\,{\sc ii}]}: ${\rm EW}=10 \rm\AA$,
corresponding to the ionizing continuum emission \citep{w99}.
Thus, we can use relation (\ref{lbolopt}) to estimate the
bolometric luminosity of the disks in BL Lac objects from their
narrow-line luminosity $L_{[\rm O\,{\rm\sc II}]}$. However, the
situation becomes quite different for those sources in which ADAFs
are present, because the spectral energy distributions of ADAFs
are significantly different from that of standard accretion disks
\citep{m97}. \citet{cao02a} have calculated the maximal optical
luminosity of BL Lac objects for the pure ADAF case, and the lower
limits on the central black hole masses of BL Lac objects are
derived from their broad-line emission. Here, we consider a more
general case, i.e., an ADAF is present near the black hole and it
transits to a cold standard disk (SD) beyond the transition radius
$r_{\rm tr}$ \citep{e97}.

The flux due to viscous dissipation in the outer region of the
disk is \be F_{\rm vis}(R)\simeq {\frac {3GM_{\rm bh}\dot M}{8\pi
R^3}}, \label{fvis} \ee which is a good approximation for $R_{\rm
tr}\gg R_{\rm in}$. The local disk temperature of the thin cold
disk is \be T_{\rm disk}(R)={\frac {F_{\rm
vis}(R)^{1/4}}{\sigma_{\rm B}^{1/4}}}, \label{tdisk} \ee by
assuming local blackbody emission. In order to calculate the disk
spectrum, we include an empirical color correction for the disk
thermal emission as a function of radius. The correction has the
form \citep{chiang02} \be f_{\rm col}(T_{\rm disk}) = f_\infty -
\frac{(f_\infty - 1) (1 +
                     \exp(-\nu_{\rm b}/\Delta\nu))} { 1 +
                     \exp((\nu_{\rm p} -\nu_{\rm b})/\Delta\nu)}, \label{fcol}
\ee
where $\nu_p \equiv 2.82k_B T_{\rm disk}/h$ is the peak frequency of a
blackbody with temperature $T_{\rm disk}$.  This expression for $f_{\rm
col}$ goes from unity at low temperatures to $f_\infty$ at high
temperatures with a transition at $\nu_{\rm b} \approx \nu_{\rm p}$.
\citet{chiang02} found  that
$f_\infty = 2.3$ and $\nu_b = \Delta\nu = 5\times 10^{15}$\,Hz do a
reasonable job of reproducing the model disk spectra of \citet{h01}.
The disk spectra can therefore be calculated by
\be
L_\nu =8\pi^2 \left( {\frac {GM}{c^2}} \right)^2
{\frac{h \nu^3}{c^2 f_{\rm col}^4} }
     \int\limits_{r_{\rm tr}}^\infty
     {\frac{r dr}{\exp(h\nu/f_{\rm col} k_B T_{\rm disk}) - 1}}.
\ee In this ADAF$+$SD scenario, the ionizing luminosity from the
disk is a combination of the emission from the inner ADAF and
outer standard disk regions. \citet{cao02a}'s calculations
indicate that the optical continuum emission from the inner ADAF
region can be neglected compared with that from outer SD region if
$r_{\rm tr}$ is around tens to several hundreds, which is due to
the low radiation efficiency of ADAFs.

\section{Results}

In Fig. \ref{fig1}, we plot the relation between the ratios
$L_{\rm bol}/L_{\rm Edd}$ and $Q_{\rm jet}/L_{\rm bol}$. For a
standard thin disk, The optical luminosity of BL Lac objects is
estimated from their narrow-line luminosity $L_{\rm [O\,{\rm \sc
II}]}$ by assuming ${\rm EW}=10{\rm \AA}$. We convert the optical
luminosity to bolometric luminosity using the relation
(\ref{lbolopt}). The black hole masses of BL Lac objects are
estimated by using the relation (\ref{mrmbh}) from the host galaxy
luminosity. The ratio  $L_{\rm bol}/L_{\rm Edd}$ is then
available. The jet power of BL Lac objects is estimated from the
low-frequency extended radio luminosity by using the relation
(\ref{qjetrad}).

For a standard thin disk, we can calculate the maximal jet power
$L_{\rm BP}^{\rm max}$ extracted from the magnetized accretion
disk as functions of black hole mass $m$ and accretion rate $\dot
m$ using Eqs. (\ref{lbpsd}), (\ref{mfpd}), and (\ref{mfdyn}). In a
similar way, the maximal jet power $L_{\rm BZ}^{\rm max}$
extracted from a spinning hole is also calculated  from Eqs.
(\ref{mfpd}), (\ref{mfdyn}), and (\ref{lbzsd}), if the black hole
mass $m$ and dimensionless black hole angular momentum $a$ are
specified. The ratios $L_{\rm bol}/L_{\rm Edd}$ and $Q_{\rm
jet}/L_{\rm bol}$ are functions of the accretion rate $\dot{ m}$
and black hole spin $a$. We vary accretion rate $\dot{m}$ and then
the relations of $L_{\rm BP}^{\rm max}/L_{\rm bol}$ and $L_{\rm
BZ}^{\rm max}/L_{\rm bol}$ with $L_{\rm bol}/L_{\rm Edd}$ are
calculated for $a=0.95$ (see Fig. \ref{fig1}). We find that the
jet power of most sources in our sample is above the maximal jet
power expected to be extracted from a magnetized accretion disk
(above the solid line in Fig. \ref{fig1}). All sources have jet
power much higher than the maximal jet power extracted by the
Blandford-Znajek mechanism (above the dotted line). This implies
that only the jets in three sources: $0828+493$, $0851+202$, and
$2240-260$ (labelled squares in Fig. \ref{fig1}, hereafter
referred as square sources) may be fuelled by the magnetized thin
disks, and no jet in the sources of this sample can be powered
only by the Blandford-Znajek mechanism, if the standard thin
accretion disks are present in these sources.

We plot the relation of jet power $Q_{\rm jet}$ with black hole mass
$M_{\rm bh}$ in Fig. \ref{fig2}. All sources have jet power less than
$0.01~L_{\rm Edd}$.
The exact value of the critical accretion rate $\dot m_{\rm crit}$ is still
unclear for an ADAF, which depends on the value of the disk viscosity parameter
$\alpha$. For AGNs, the value of the critical accretion rate $\dot
m_{\rm crit}$ is probably around $\sim 0.01$ \citep{n02}.
The maximal jet power extracted from a rapidly
spinning black hole $a=0.95$ is calculated as a function of black hole
mass $M_{\rm bh}$ assuming  $\dot{m}_{\rm crit}=0.01$.

The relation between black hole mass $M_{\rm bh}$ and the central
optical ionizing continuum luminosity $L_{\lambda}$ at
3727~$\rm\AA$ is plotted in Fig. \ref{fig3}. The optical continuum
emission from a pure ADAF can be calculated by using the approach
proposed by \citet{m97}. We use the same approach proposed by
\citet{cao02a}  to calculate the maximal optical continuum
emission from an ADAF as a function of black hole mass $M_{\rm
bh}$. The maximal optical continuum emission requires the
parameter $\beta=0.5$, which describes the magnetic field strength
with respect to gas pressure, and viscosity $\alpha=1$. Changing
the value of accretion rate $\dot{m}$, we can find the maximal
optical continuum luminosity for given black hole mass (see Cao,
2002a for details). It is found that all sources have optical
ionizing luminosity higher than the maximal optical luminosity
expected from pure ADAFs, which implies that the emission from
pure ADAFs is unable to explain the optical ionizing luminosity of
these sources.  The optical spectra of the standard accretion
disks are also calculated for different accretion rates
$\dot{m}=0.001$, 0.01, and $0.1$, respectively. Most sources,
except three square sources, have optical ionizing luminosity
below the line of $\dot{m}=0.01$ expected from the standard disk.
We further calculate the optical spectra of ADAF$+$SD systems as
described in Sect. 6, for different values of transition radius
$r_{\rm tr}$. All these sources lie in the region between $r_{\rm
tr}=40$ and $150$ for $\dot{m}=0.01$. If $\dot{m}=0.1$ is adopted,
the transition radii of the disks are required to be in the range
of $100-400$ to explain the optical ionizing emission.

In Figs. \ref{fig4} and \ref{fig5}, we present the results as that
in Figs. \ref{fig1} and \ref{fig2}, but for $f=10$ is adopted in
the normaliztion of the jet power estimate (\ref{qjetrad}).

\section{Discussion}

For a standard magnetized accretion disk, the strength of the
large-scale ordered field of the disk is related with the
scale-height of the disk. The scale-height of the disk is
proportional to dimensionless accretion rate $\dot{m}$. Thus, the
maximal jet power extracted from the disk or the spinning black
hole depends on the accretion rate $\dot{m}$. We found that only
three sources $0828+493$, $0851+202$, and $2240-260$ have lower
jet power than the maximal jet power expected by the
Blandford-Payne mechanism for a standard accretion disk. This may
imply that the accretion disks in most BL Lac objects  (except the
three sources) are not standard thin disks.

If ADAFs are present in the BL Lac objects, there is a maximal jet
power for a given black hole mass due to an upper limit on the
accretion rate $\dot{m}_{\rm crit}$ for an ADAF \citep{ny95}.
Compared with the standard disk case, ADAFs have much stronger
magnetic field strength \citep{l99,an99}. We found in Fig.
\ref{fig2} that the power of jets in all sources is less than
$0.01~L_{\rm Edd}$. The power of the jets in most sources can be
explained by the Blandford-Znajek mechanism if an ADAF is
surrounding a rapidly spinning black hole at an accretion rate
$\dot{m}=0.01$, while six sources with jet power higher than the
maximal jet power in the case of $a=0.95$. The jets in these six
sources may be powered by the Blandford-Payne mechanism from the
ADAFs (e.g. ADIOS proposed by Blandford \& Begelmann 1999).
Another possibility may be that the critical accretion rate
$\dot{m}_{\rm crit}$ is higher than $0.01$. If the critical
accretion rate is $0.1$, the power of jets in all sources can be
explained in the frame of the Blandford-Znajek mechanism for ADAFs
surrounding rapidly spinning black holes.

\begin{figure}
\plotone{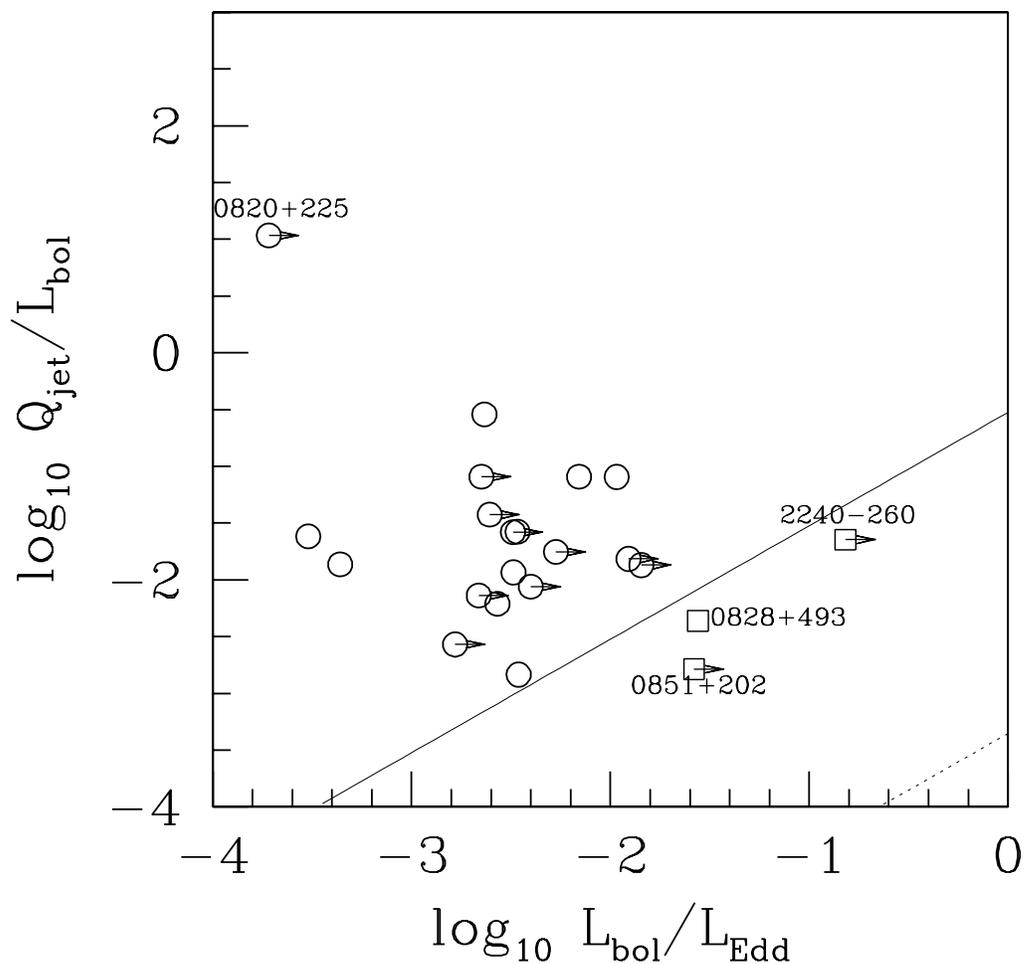} \caption{The ratio $L_{\rm bol}/L_{\rm Edd}$
versus the ratio $Q_{\rm jet}/L_{\rm bol}$ ($f=1$ is adopted). The
solid line represents the maximal jet power $L_{\rm BP}^{\rm max}$
extracted from a standard accretion disk (the Blandford-Payne
mechanism), while the dotted line represents the maximal jet power
$L_{\rm BZ}^{\rm max}$ extracted from a rapidly spinning black
hole $a=0.95$. The sources below the solid line are labelled as
squares. \label{fig1}}
\end{figure}

\begin{figure}
\plotone{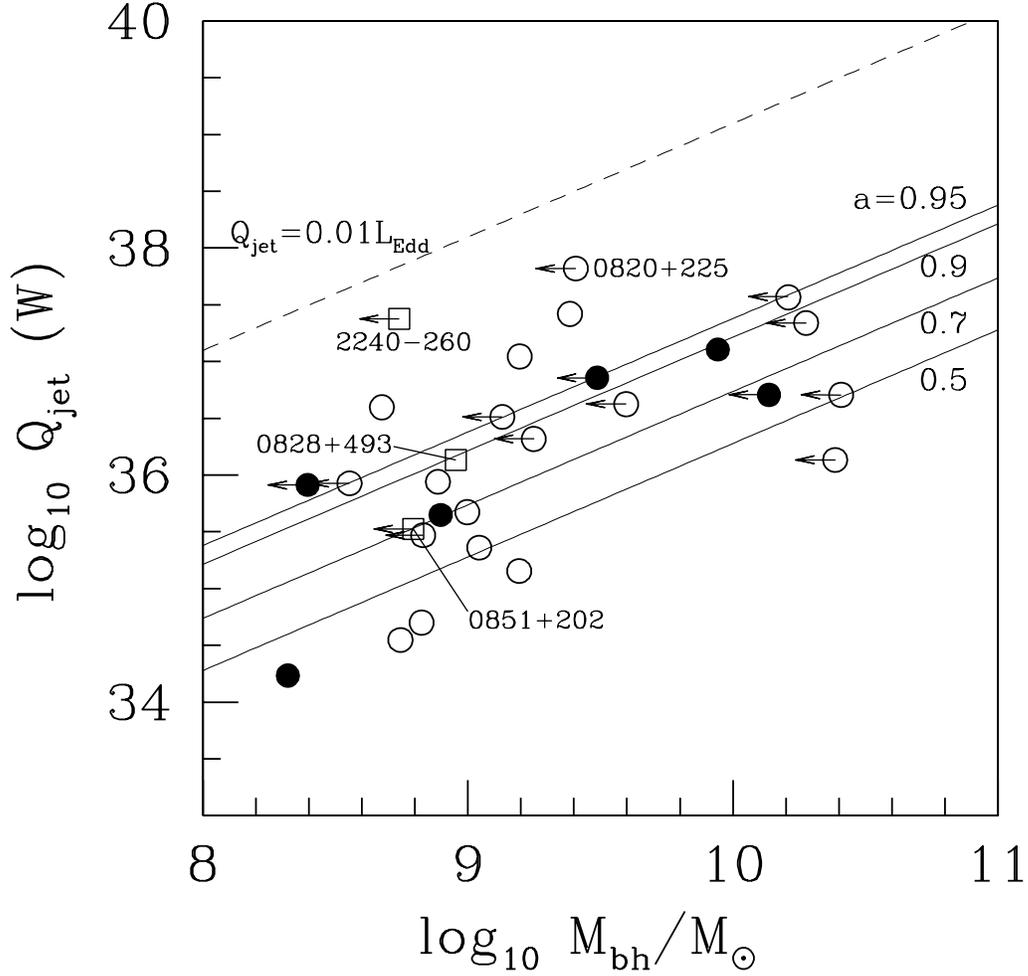} \caption{ The black hole mass $M_{\rm bh}$
versus jet power $Q_{\rm jet}$($f=1$ is adopted). The full circles
represent the sources without emission line data. The solid lines
represent the maximal jet power $L_{\rm BZ}^{\rm max}$ extracted
from the spinning black holes surrounded by ADAFs as functions of
$M_{\rm bh}$ for different values of spin $a$, respectively
($\dot{m}_{\rm crit}=0.01$ is adopted). The dashed line represents
the jet power $Q_{\rm jet}=0.01L_{\rm Edd}$. \label{fig2}}
\end{figure}

\begin{figure}
\plotone{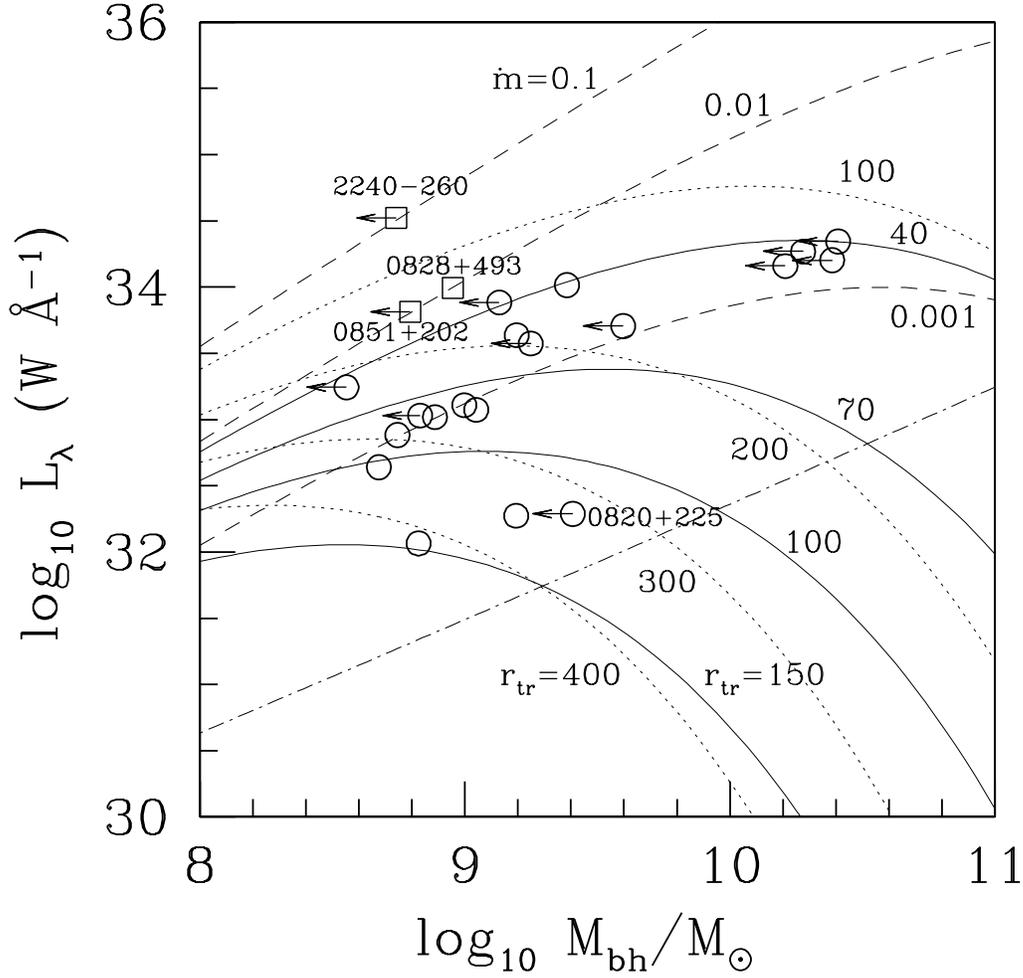} \caption{ The black hole mass $M_{\rm bh}$
versus optical luminosity $L_{\lambda}$ at 3727~$\rm \AA$. The
solid lines represent for ADAF$+$SD models with different
transition radii $r_{\rm tr}$ for $\dot{m}=0.01$, while the dotted
lines represent the cases of $\dot{m}=0.1$. The dashed lines
represent standard accretion disks with $\dot{m}=0.001$, 0.01, and
0.1, respectively. The dash-dotted line is the maximal optical
luminosity as a function of black hole mass $M_{\rm bh}$ for the
pure ADAF case. \label{fig3}}
\end{figure}

For given black hole mass, there is an upper limit on the optical
continuum emission from an ADAF \citep{cao02a}. As the radiative
efficiency of ADAF is very low, even the maximal optical continuum
luminosity of an ADAF is much fainter  than the standard accretion
disk at the same accretion rate. We test ADAF scenario in Fig.
\ref{fig3}, and found that the pure ADAF model is unable to
produce sufficient optical ionizing emission derived from
{[O\,{\sc ii}]}-line emission for all sources (see the dash-dotted
line in Fig. 3). The optical continuum emission from three sources
$0828+493$, $0851+202$, and $2240-260$ can be explained as
emission from standard disks accreting at $\dot{m}=0.01$ to $0.1$
respectively, which is consistent with the fact that only the jets
in these three sources may be produced by the Blandford-Payne
mechanism for standard disks (see Fig. \ref{fig1}). For the
remainder, we propose their optical continuum emission to be
produced by the ADAF$+$SD systems. The optical continuum
luminosity of the ADAF$+$SD system is available as a function of
black hole mass $m$, while the accretion rate $\dot m$ and the
transition radius $r_{\rm tr}$ are specified. Such calculations
presented in Fig. \ref{fig3} show that the optical ionizing
continuum emission from these BL Lac objects can be explained by
the ADAF$+$SD model, if $r_{\rm tr}$ are in the range of $40$ to
$150$ for $\dot{m}=0.01$. We also calculate the cases of
$\dot{m}=0.1$, though it is still unclear whether an ADAF can
exist in the inner region of the disk for such a high accretion
rate. We find that the transition radii are in the range of
$100-400$ for $\dot{m}=0.1$. The jets in these BL Lac objects may
be accelerated by either the Blandford-Znajek mechanism or the
Blandford-Payne mechanism (or by both two mechanisms) from the
inner ADAF region, while the optical ionizing continuum emission
is mainly from the outer standard disk region.

For adiabatic inflow-outflow solutions (ADIOSs) \citep{bb99}, the
accretion rate of the disk is a function of radius $r$ instead of
a constant accretion rate along $r$ for a pure ADAF. The ADIOS is
described by the similar equations for an ADAF, while a
$r$-dependent accretion rate $\dot{m}(r)$ is adopted instead. The
accretion rate at the inner edge of the disk for an ADIOS should
be at least as low as that required for pure ADAF solutions to
keep the flow advection dominated. ADIOS has similar structure and
a similar upper limit on the accretion rate at the inner edge of
the disk as that of a pure ADAF, if the wind is not strong
\citep{ccy02}. In the case of strong winds, the gas swallowed by
the black hole for an ADIOS is only a small fraction of the rate
at which it is supplied, as most of the gas is carried away in the
wind before it reaches the black hole. The maximal accretion rate
of an ADIOS at its inner edge is significantly lower than that for
a pure ADAF, if a strong wind is present. The maximal jet power
extracted from a spinning black hole by the Blandford-Znajek
mechanism for ADIOS cases should therefore be lower than that for
pure ADAF cases. So, the power extracted from rapidly spinning
black holes for ADIOS cases is insufficient for strong jets in
some BL Lac objects.

Recently, \citet{y01} proposed that a hot luminous disk may be
present if the accretion rate is higher than the critical value
$\dot{m}_{\rm crit}$ of the ADAF.  This hot luminous disk may
exist even if the accretion rate $\dot{m}$ is as high as unit. The
disk is much luminous than a pure ADAF due to its higher accretion
rate and accretion efficiency. Thus, the maximal jet power $L_{\rm
BZ}^{\rm max}$ extracted from a spinning black hole surrounded by
such a hot luminous disk may be higher than that for a pure ADAF.
This may be helpful to account for high power of jets in some BL
Lac objects. The hot luminous disk has lower accretion efficiency
than the standard thin disk. The optical continuum emission of
these sources can be mainly from the standard disk region, if the
hot luminous disk transits to a standard disk in the outer region
of the disk. Such a hot luminous disk+SD model may work in some BL
Lac objects.

\begin{figure}
\plotone{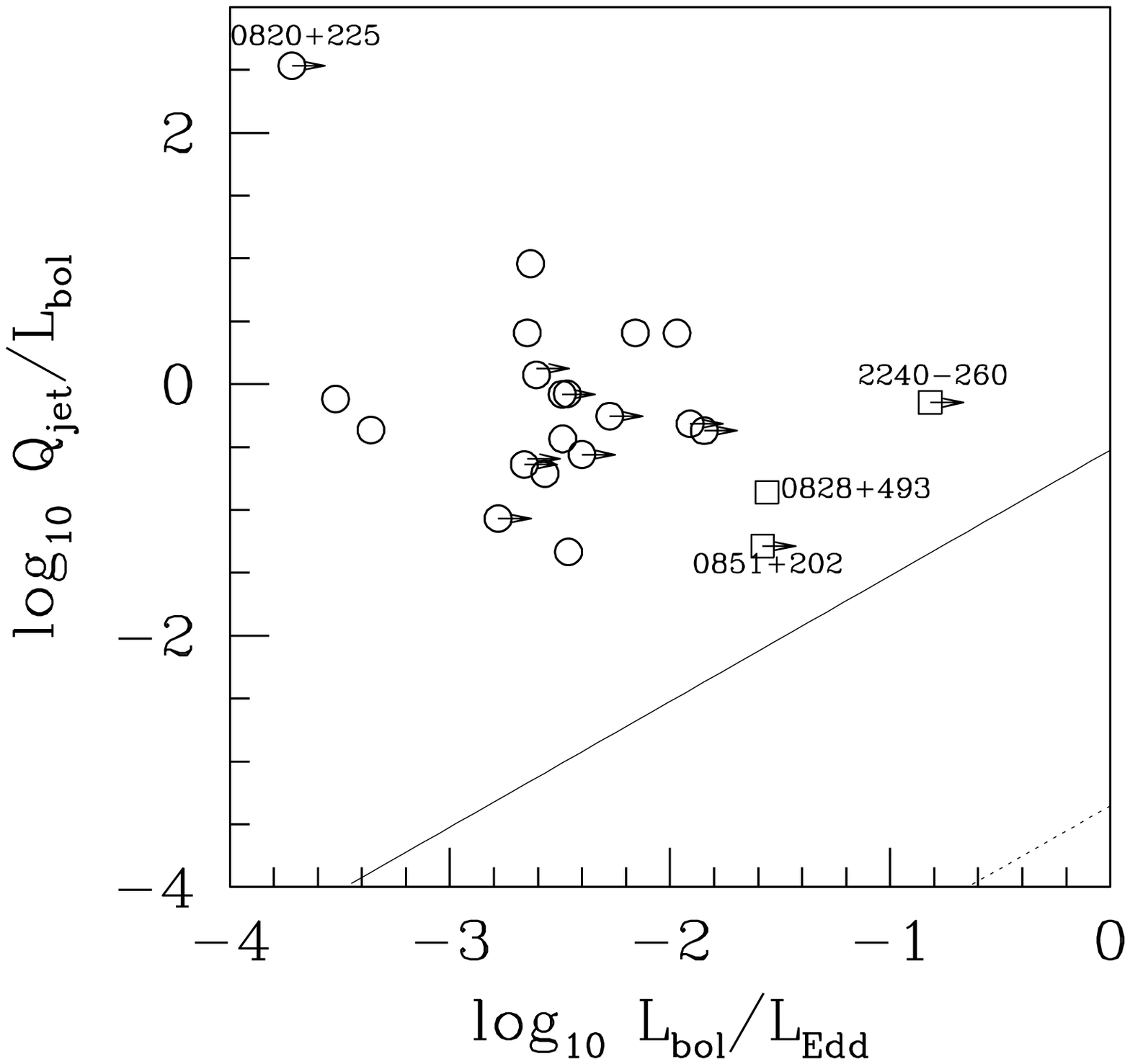} \caption{
Same as Fig. \ref{fig1}, but $f=10$ is adopted.
\label{fig4}}
\end{figure}

\begin{figure}
\plotone{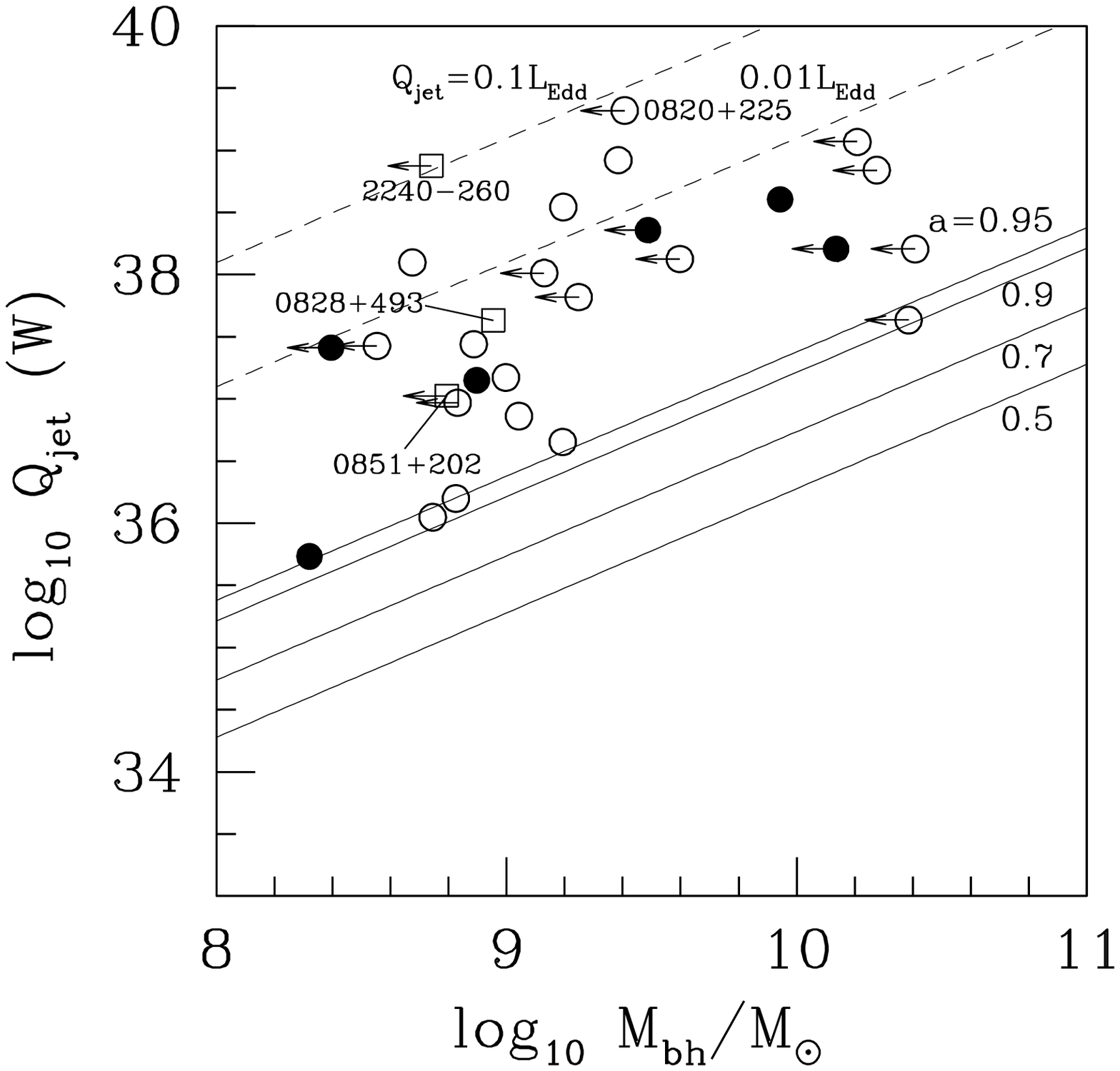} \caption{
Same as Fig. \ref{fig2}, but $f=10$ is adopted.
\label{fig5}}
\end{figure}

We have adopted $f=1$ in the normalization of the jet power
estimate. In order to explore the case of $f$ greater than unit,
we perform calculations for $f=10$ in Figs. \ref{fig4} and
\ref{fig5}. In this case, the jets in all sources cannot be
accelerated by the standard magnetized disks. Only the
Blandford-Payne mechanism for ADAFs (or ADIOSs) accreting at high
rates around $\sim0.1$ are able to produce strong jets in some
sources. Even if this is the case, standard disks in the outer
regions of the disks are still required to produce their optical
ionizing luminosity (see Fig. \ref{fig3}).

In this paper, the accretion rate $\dot{m}$ is adopted as a free
parameter to compare our calculations with observations. For an
accretion rate $\dot{m}$ being less than $0.01$ is adopted, for
example $\dot{m}=0.001$, we find that pure standard accretion
disks are required to explain their optical continuum emission at
least for some sources, i.e., the sources locate between the lines
of $\dot{m}=0.001$ and $0.01$ in Fig. \ref{fig3}. However, the
standard magnetized thin disks are not able to fuel the jets in
these sources (see Fig. \ref{fig1} for $f=1$ and Fig. \ref{fig4}
for $f=10$). This indicates that the accretion rates $\dot{m}$, at
least in these sources, should be around $\sim0.001-0.01$. It
means that the inner region of the disk would be advection
dominated if $\dot{m}\sim 0.001-0.01$, which may imply that the
critical accretion rate $\dot{m}_{\rm crit}$ is probably less than
$\sim 0.001-0.01$ at least for these sources. The ADAF$+$SD
scenario presented in this work is in general consistent with the
models recently proposed  for disk-jet connection in AGNs
\citep{h03,l03,falcke03}.

The transition radius $r_{\rm tr}$ is another parameter adopted in
our calculations of the continuum emission from the disks. One
possible mechanism for the transition might be the evaporation of
the disk \citep{mm94,liu99,rc00,mm01}. In this model, the
disk-corona structure gradually changes to a pure vertically
extended coronal or ADAF in the inner region of the disk. In the
outer region of the disk, the cold disk and the hot corona above
are in pressure equilibrium. In this scenario, most gravitational
energy of the accretion matter is released in the hot corona
\citep{hm91,km94,sz94}. A small fraction of the soft photons from
the cold disk is Compton up-scattered to X-ray photons by the hot
electrons in the corona. Roughly about half of the scattered X-ray
photons illuminate the cold disk \citep{hm91,no93,c98,k01}. Most
soft photons from the disk leave the system without being
scattered in the corona, and they are observed as optical/UV
continuum.  The cold disk in the disk-corona system has roughly
about half brightness of a standard disk without a corona, if they
are accreting at the same rate \citep{hm91}. The optical emission
is therefore mainly from the cold disk, and the contribution to
the optical emission form the corona and ADAF in the inner region
of the disk can be neglected. If the disk-corona structure extends
to the inner edge of the disk, the cold disk in the disk-corona is
too bright to explain the ionizing luminosity of most BL Lac
objects in this sample.

\acknowledgments
I thank the referee for his helpful comments and suggestions.
This work is supported by NSFC(No. 10173016) and the NKBRSF (No.
G1999075403). This research has made use of the NASA/IPAC Extragalactic
Database (NED), which is operated by the Jet
Propulsion Laboratory, California Institute of Technology,
under contract with the National Aeronautic and Space Administration.

\clearpage

\begin{deluxetable}{ccccccccc}
\tabletypesize{\scriptsize}
\tablecaption{Data of BL Lac objects}
\tablewidth{0pt}
\tablehead{
\colhead{Source} &
\colhead{Redshift}   &
\colhead{$m_R$(host)}   &
\colhead{$\log~L_{\rm [O~II]}$} &
\colhead{References} &
\colhead{$\log~L_{\rm ext,151}$  }  &
\colhead{References} &
\colhead{$\log~M_{\rm bh}/M_\odot$ }  &
\colhead{$\log~Q_{\rm jet}$  }  }
\startdata
 0118$-$272 &  0.559 & $>$17.96 &  ...  & ...  & 27.21 & C99 & ~$<$9.49 & 36.86 \\
 0138$-$097 &  0.733 & $>$18.38 & 34.71 & RS01 & 26.94 & C99 & ~$<$9.60 & 36.62 \\
 0235$+$164 &  0.940 & $>$17.35 & 35.35 & S93 & 27.03 & C99 & $<$10.41 & 36.71 \\
 0426$-$380 &  1.030 & ~~18.50 &  ...  & ... & 27.50 & C99 &  ~~~9.94 & 37.11 \\
 0521$-$365 &  0.055 & ~~14.35 & 33.64 & T93 & 26.91 & AU85 & ~~~8.68 & 36.60 \\
 0537$-$441 &  0.896 & $>$17.50 & ~~35.27$^{\rm a}$ & S93 & 27.77 & C99 & $<$10.28 & 37.34 \\
 0735$+$178 &  0.424 & $>$19.50 &  ...  & ... & 26.11 & C99 & ~$<$8.40 & 35.91 \\
 0754$+$100 &  0.670 & $>$17.09 &  ...  & ... & 27.03 & M93 & $<$10.14 & 36.71 \\
 0820$+$225 &  0.951 & $>$19.38 & ~~33.29$^{\rm a}$ & S93 & 28.33 & C99 & ~$<$9.41 & 37.82 \\
 0823$+$033 &  0.506 & $>$19.04 & 34.03 & S93 & 25.59 & C99 & ~$<$8.83 & 35.47 \\
 0828$+$493 &  0.548 & ~~18.98 & 34.99 & RS01 & 26.36 & C99 & ~~~8.95 & 36.13 \\
 0829$+$046 &  0.180 & ~~16.54 &  ...  & ... & 25.80 & AU85 & ~~~8.90 & 35.65 \\
 0851$+$202 &  0.306 & $>$17.95 & ~~34.81$^{\rm b}$ & S89 & 25.66 & C99 & ~$<$8.79 & 35.53 \\
 0954$+$658 &  0.367 & $>$18.85 & 34.24 & RS01 & 26.12 & C99 & ~$<$8.55 & 35.93 \\
 1144$-$379 &  1.048 & $>$19.93 & ~~34.57$^{\rm a}$ & S89 & 26.58 & C99 &  ~$<$9.25 & 36.32 \\
 1418$+$546 &  0.152 & ~~15.87 & 34.07 & S93 & 25.46 & C99 &  ~~~9.04 & 35.36 \\
 1538$+$149 &  0.605 & ~~18.73 & 34.64 & S93 & 27.43 & C99 &  ~~~9.20 & 37.04 \\
 1749$+$096 &  0.320 & ~~18.15 & 33.88 & S93 & 24.52 & M93 &  ~~~8.75 & 34.55 \\
 1749$+$701 &  0.770 & $>$16.92 & 35.20 & S89 & 26.37 & C99 & $<$10.39 & 36.13 \\
 1803$+$784 &  0.684 & $>$19.15 & ~~34.88$^{\rm b}$ & L96 & 26.81 & C99 &  ~$<$9.13 & 36.51 \\
 1807$+$698 &  0.051 & ~~13.54 & 34.11 & S93 & 25.83 & C99 &  ~~~9.00 & 35.67 \\
 1823$+$568 &  0.664 & ~~18.57 & 35.02 & S93 & 27.87 & C99 &  ~~~9.39 & 37.42 \\
 2007$+$777 &  0.342 & ~~18.02 & 34.02 & S89 & 26.14 & C99 &  ~~~8.89 & 35.94 \\
 2131$-$021 &  1.285 & $>$18.50 & 35.16 & RS01 & 28.04 & C99 & $<$10.21 & 37.57 \\
 2200$+$420 &  0.069 & ~~14.55 & ~~33.07$^{\rm b}$ & S93 & 24.69 & C99 &  ~~~8.83 & 34.70 \\
 2201$+$044 &  0.027 & ~~13.50 &  ...  & ... & 24.15 & A98 &  ~~~8.32 & 34.23 \\
 2240$-$260 &  0.774 & $>$20.22 & 35.52 & S93 & 27.81 & C99 &  ~$<$8.74 & 37.38 \\
 2254$+$074 &  0.190 & ~~16.07 & ~~33.27$^{\rm b}$ & S93 & 25.22 & C99 & ~~~9.19 & 35.15 \\
 \enddata
 \tablenotetext{a}{converted from Mg\,{\sc ii}}
 \tablenotetext{b}{converted from [O\,{\sc iii}].}
\tablecomments{Column (1): source name; Column (2): redshift;
Column (3): $R$-band magnitude of the host galaxy; Column (4):
narrow-line luminosity $L_{\rm [O\,{\rm \sc II}]}~({\rm W})$;
Column (5): references
for $L_{\rm [O\,{\rm \sc II}]}$; Column (6): extended radio luminosity at 151~MHz in
the rest frame of the source (${\rm W~Hz}^{-1}$); Column (7):
references for $L_{\rm ext,151}$; Column (8): black hole mass;
Column (9): jet power (W). }
\tablerefs{AU85: \citet{au85}; C99: \citet{c99}; L96: \citet{l96};
M93: \citet{m93}; RS01: \citet{rs01}; S89: \citet{s89};
S93: \citet{s93}; T93: \citet{t93}; }
\end{deluxetable}

\clearpage


\end{document}